\documentclass[secnumarabic,preprint]{revtex4-1}
\usepackage{amssymb,amsbsy,amsmath}
\usepackage[english]{babel}
\usepackage{graphicx}
\newcommand{\ha}{\mbox{\small$\frac{1}{2}$}}

\newcommand{\im}{\mathrm{i}\,}

\newcommand{\lab}[1]{\label{#1}}
\newcommand{\re}[1]{(\ref{#1})}
\newcommand{\nn}{\nonumber}

\newcommand{\B}[1]{\boldsymbol{#1}}
\newcommand{\s}[1]{\mathsf{#1}}

\newcommand{\BOm}{\boldsymbol{\Omega}}

\newcommand{\inta}{\hspace*{-.2em}\int \hspace*{-.2em}}
\newcommand{\intab}{\hspace*{-.2em}\int \hspace*{-.4em}\int \hspace{-.2em}}
\newcommand{\ints}{\lefteqn{\hspace*{-.2em}\int \hspace*{-.8em}\int
\hspace{-.3em}}\hspace{.1em}\diagdown\,\,\,}
\newcommand{\D}[2]{{\rm d}^{#1}{#2}\,}
\newcommand{\cc}[1]{\lefteqn{\stackrel{*}{\phantom{#1}}}#1}

\newcommand{\lr}[1]{\sqrt{\vphantom{\dot x^2_a}\dot x^2_{#1}}}
\begin{document}

%\preprint{}

\title{Quantization of almost-circular orbits in the Fokker action formalism.
Regge trajectories.}
\author{Askold Duviryak}
\email[]{duviryak@icmp.lviv.ua}
\affiliation{Department for Computer Simulations of Many-Particle Systems,
Institute for Condensed Matter
Physics of NAS of Ukraine, Lviv, UA-79011, Ukraine}

\date{\today}

\begin{abstract}
A relativistic quark model of mesons formulated within the formalism
of Fokker-type action integrals is proposed, in which an interquark
interaction is mediated by scalar-vector superposition of higher
derivative fields. In the non-relativistic limit the model describes
a two-particle system with the linear potential. In order to analyze
the model in the essentially relativistic domain the perturbed
circular orbit approximation and certain principle of selection of
physically meaningful solutions are applied which permit one to perform
the canonical quantization of the model. It is shown that the model
reproduces well specific features of the light meson spectroscopy.
\end{abstract}

\maketitle                   % Produces the title.

%%%%%%%%%%%%%%%%% Section 1 %%%%%%%%%%%%%%%%%%%%%%%%%%%%%

\section{Introduction}

It is known that spectra of heavy mesons (contaning c and b quarks)
are described well by means of potential models
with the non-relativistic Cornell potential $u(r) = u_0 -\alpha/r + ar$
and various quasi-relativistic corrections of scalar-vector type
\cite{Eich75,LSG91,H-L92,H-M02}. The potential is QCD-motivated:
its Coulomb part is a
non-relativistic limit of the one-gluon exchange interaction while the
linear part comes from the Wilson loop. The latter is also related to
a string conception of hadrons \cite{Nie77,J-N79,Sim94}. Constants $u_0$,
$\alpha$ and $a$ vary from one model to another. In particular,
the {\em string tension} parameter $a$ is frequently used as
an adjustable parameter from the range $a=0.15\div0.3$~GeV$^2$
\cite{Eich75,LSG91,H-L92,H-M02} although the most conventional value
$a=0.18\div0.2$~GeV$^2$ is substantiated by QCD simulations on the lattice
\cite{B-S09}.

Mass spectra of light mesons (consisting
of u, d and s quarks) possess characteristic features which
can be summarized roughly in the following
idealized picture \cite{Sim94,B-P91,Duv06,Duv08}:
\begin{enumerate}
\item
Meson states are clustered in the family of straight lines in the
($M^2,j$)--plane known as Regge trajectories.
\item
    Regge trajectories are parallel; {\em slope} parameter $\sigma$
is an universal quantity, $\sigma=1.15\div1.2\,{\rm GeV}^2$.
\item
    As states of quark-antiquark system mesons can be classified
non-relativistically, by $\ell$ and $n_r$ (the orbital and radial
quantum numbers) as well as by $s$ and $j$ (the total spin and
angular momentum).
\item
    Spectrum is $\ell$$s$-degenerated, i.e., masses are distinguished by
$\ell$ (not by $j$ or $s$) and $n_r$.
\item
   States of different $\ell$ and $n_r$ possess an accidental
degeneracy which causes a tower structure of the spectrum.
\end{enumerate}
Items 1--4 imply that in the ($M^2,\ell$)--plane meson states form
into strait lines too: the principal ($n_r=0$) Regge trajectory
built up of the set of degenerated singlet ($s=0$) and triplet
($s=1$) states, and the family of daughter trajectories ($n_r=1,2,\dots$).
Hence energy levels of q-\=q system can be described by a formula:
%
%           Equation 1.1
\begin{equation}\lab{1.1}
M^2\approx\sigma(\ell + \varkappa n_r + \zeta),
\end{equation}
where the intercept constant $\zeta$ depends on a flavor content of
mesons ($\zeta\approx1/2$ for ($\pi$-$\rho$)--family of mesons; it
grows together with quark masses). Finally, the accident degeneracy
(the item 5) constraints the constant coefficient $\varkappa$
determining the spacing of daughter trajectories to an integer
or a rational number \cite{GCO90}.

Light mesons are essentially relativistic two-quark systems, and
considerable amount of various relativistic models has been invented
for their description. The most elegant and historically important
among them are the simple relativistic oscillator (and its variations)
\cite{K-N73,Tak79,I-O94} and string models \cite{Nie77,J-N79,Sim94}.
They lead exactly or asymptotically (at large $\ell$) to the formula
\re{1.1} with $\varkappa=2$. Besides, the string models tie the slope
and the string tension together, $\sigma=ka$, with the slope coefficient
$k=2\pi\approx6.3$, so that, the value $a=0.18$~GeV$^2$ is
preferable.

Farther relativistic potential models are based on various relativistic
generalizations of Schr\"odinger equations with a confining (Cornell or more
complicated) potential such as one- \cite{HLS94,Hay96,LRR08} and two-particle
\cite{KvA84,Saz86,S-C93,KvA04,M-N05,Duv06,Duv08} Dirac
equations etc \cite{B-R85,Khr87-1}. These models incorporate the description of
heavy and light hadrons and, in most, reveal asymptotically linear Regge
trajectories \re{1.1} of the slope $\sigma=ka$ with the slope coefficient
$k=4\div8$ and with the daughter spacing coefficient $\varkappa=2$.
With this kind of degeneracy, i.e., of  ($\ell{+}2n_r$)-type, however,
a certain number of states falls out the description \cite{Sim94,Duv06}.
The value $\varkappa=1$ is more adequate to
experimental data. In particular, it follows (in the limit $\ell\gg1$)
from the mass formulae derived by means of the Dirac-type equation model
\cite{Khr87-1} and selection rules superinduced by hands,
and used for a description of $\pi$-, $\rho$- \cite{Khr87-2}
and K-trajectories \cite{Khr89}.

In the present paper we consider the relativistic potential model
of mesons which reveals asymptotically linear  Regge trajectories with
native ($\ell{+}n_r$)-degeneracy. A classical prototype of the model
was formulated independently by Rivacoba \cite{Riv84}
and Weiss \cite{Wei86} by means of the Fokker-type action integral
\cite{Hav71,Ker72} related, in turns, to a higher-derivative
gauge field theory \cite{Duv99,L-M12}. Namely, the interaction
between particles is described
in terms of a time-symmetric Green function
of a fourth-order field equation.

Hamiltonization and quantization of Fokker-type systems
is rather challenging problem in view of a time-nonlocal
character of the interaction \cite{GKT87,JJLM87,JJLM89,L-V94}. The
Hamiltonian description in this case can be built by means of
approximated methods \cite{GKT87,JJLM89} which, in most, are not
appropriate for strongly coupled systems.

For particular time-symmetric Fokker-type systems
one can invent naturally
time-asymmetric counterparts in which a time-nonlocality
is removed \cite{Duv97}. The Rivacoba-Weiss model is the case.
For this but time-asymmetric model an exact Hamiltonian formulation
(see \cite{Duv97} for general formalism) and
the corresponding quantum description was elaborated
\cite{Duv99,Duv01}. In despite of an admired degeneracy (i.e., with
$\varkappa=1$), the slope of asymptotic Regge trajectories
turned out to be overestimated, with the
coefficient $k=3\sqrt{6}\approx10.4$. The reason perhaps
resides in the fact that the vector character of
interaction brought into the model from the underlying
gauge theory is not quite suited to an actual nature of a
confining interaction in hadrons. Unfortunately,
the Fokker-type model of scalar confinement without
time-nonlocality is unknown.

Recently, a quantization method of two-particle Fokker-type systems
in an almost-circular-orbit (ACO) approximation has been proposed
by the author \cite{Duv12a}.
The method is appropriate for strongly coupled systems.
Here it is applied to a quantization of the
time-symmetric Rivacoba-Weiss model.
Moreover, the analogue of the Rivacoba-Weiss
model with scalar confining interaction is built,
and the scalar-vector superposition model is
considered. It is studied an asymptotic behavior of the Regge
trajectories, from which the slope and daughter spacing
coefficients are found and compared with data from experiment
and other potential models.

%%%%%%%%%%%%%%%%% Section 2 %%%%%%%%%%%%%%%%%%%%%%%%%%%%%

\section{Various formulations of Fokker-type action integral
with a vector linear confinement.}

We start with the manifestly covariant two-particle action
%
%           Equation 2.1
\begin{equation}\lab{2.1}
I = I_{\rm free}+I_{\rm int} \qquad\mbox{where}\quad\ I_{\rm free}
\!= -\!\sum\limits_{a=1}^{2} m_a\inta \D{}{\tau_a} \lr a,
\end{equation}
and $I_{\rm int}$ is the Fokker action integral
\cite{Hav71,Ker72} describing an interaction. For the arbitrary
interaction of a vector type we have:
%
%           Equation 2.2
\begin{equation}\lab{2.2}
I_{\rm int}^{\rm(v)} = -\intab \D{}{\tau_1}\D{}{\tau_2}\,
\dot x_1\!\cdot\!\dot x_2\, G(x_{12}^2).
\end{equation}
In eqs. \re{2.1} and \re{2.2} $m_a$ is a rest mass of $a$th particle ($a{=}1,2$);
$x^{\mu}_a(\tau_a)$ ($\mu{=}\overline{0,3}$) are covariant coordinates of a world line of
$a$th particle parameterized by an arbitrary evolution parameter
$\tau_a$; $\dot x^{\mu}_a(\tau_a) \equiv
dx^{\mu}_a/d\tau_a$; $x_{12}^\mu\equiv x_1^\mu(\tau_1)-x_2^\mu(\tau_2)$;
$x_{12}^2\equiv\eta_{\mu\nu}x_{12}^\mu x_{12}^\nu$;
the function $G(x_{12}^2)$ is usually proportional to a symmetric Green function of
an appropriate field equation, or it may be chosen phenomenologically.
We use the time-like Minkowski metrics, {\em i.e.},
$\parallel\!\eta_{\mu\nu}\!\parallel~=~\mathrm{diag} (+,-,-,-)$,
and put the light speed to be unit, $c=1$.

If one chooses
$G(x_{12}^2)\propto\delta(x_{12}^2)$ where $\delta(x^2)$ is the symmetric Green function
of the d'Alembert equation $\square\delta(x^2)=4\pi\delta(x)$, one arrives at
the Wheeler-Feynman electrodynamics \cite{W-F49}.

Let us consider the Fokker-type action proposed by Weiss \cite{Wei86}. It corresponds to
the choice $G(x^2)\propto\Theta(x^2)$ in \re{2.2} (where $\Theta(x)$ is the Heaviside
step function) with some coefficient of proportionality which we specify here as follows:
%
%           Equation 2.3
\begin{equation}\lab{2.3}
G(x^2)=-\ha a\Theta(x^2),\qquad a>0.
\end{equation}
In the non-relativistic limit the Weiss action
leads \cite{Gai82} to the interaction potential:
%
%           Equation 2.4
\begin{equation}\lab{2.4}
U(r)=\int^{\infty}_{-\infty}\D{}\vartheta G(\vartheta^2-r^2)
=-a\int^{\infty}_r\D{}\vartheta = a(r-\infty)
\end{equation}
which corresponds to a linear confinement up to unessential infinite constant.

As it is shown in \cite{Duv99,L-M12} the Weiss action principle is related
to the higher-derivative theory of the vector field proposed by Kiskis \cite{Kis75}
and to its later non-Abelian version \cite{AAB82,A-A84}. In particular, the function
$\Theta(x^2)$ is a symmetric fundamental solution of the equation:
%
%           Equation 2.5
\begin{equation}\lab{2.5}
\square^2\Theta(x^2)=16\pi\delta(x).
\end{equation}
The Fourier transform of this solution $\propto1/k^4$  coincides with
the infrared asymptotics of gluon propagator \cite{AAB82}.

An infinite constant in r.-h.s. of \re{2.4} indicates that the
Fokker action integral \re{2.2} with the Green function \re{2.3} is
not well posed from the mathematical viewpoint. A formal causal
structure of the interaction is that as if each point (say, $x_a$)
of a world line of one particle is related to infinite segments of
another word line lying inside the light cone with the center $x_a$,
and the contribution of these segments in the action is infinite;
see Fig.1(a). Physically it is not crucial since a variation of the
action \re{2.1}-\re{2.3} turns $\Theta(x^2)$ into its derivative
$\Theta'(x^2)=\delta(x^2)$, and Euler-Lagrange equations relate
points of particle world lines along generatrices of light cones only;
see Fig.1(b).
Nevertheless, integrals of motion such as the energy and the angular
momentum turns out divergent. In order to avoid this difficulty one
can reformulate the Fokker action \re{2.2}, \re{2.3} via the
integration by parts \cite{Kat69}:
%
%           Equation 2.6
\begin{eqnarray}\lab{2.6}
I_{\rm int}^{\rm(v)} &=& \frac a2\int^{\infty}_{-\infty}\int^{\infty}_{-\infty}
\D{}{\tau_1}\D{}{\tau_2}\,\dot x_1\!\cdot\!\dot x_2\, \Theta(x_{12}^2)\nn\\
&=&-a\int^{\infty}_{-\infty}\int^{\infty}_{-\infty}\D{}{\tau_1}\D{}{\tau_2}\,
(x_{12}\!\cdot\!\dot x_1)(x_{12}\!\cdot\!\dot x_2)\,\delta(x_{12}^2)\\
&&{}-\frac a4\Theta(x_{12}^2)x_{12}^2\Big|_{\tau_1=-\infty}^{\tau_1=\infty}\Big|_{\tau_2=-\infty}^{\tau_2=\infty}.
\nn
\end{eqnarray}
The last divergent term does not contribute in the equations of
motion, and we arrive at the equivalent formulation of the problem
proposed earlier by Rivacoba \cite{Riv84}. The Fokker-type integral
\re{2.6} itself describes an interaction with the causal structure
of Fig.1(b), as in the Wheeler-Feynman electrodynamics, and leads to
finite integrals of motions.
\begin{figure}[h]
\includegraphics[scale=1]{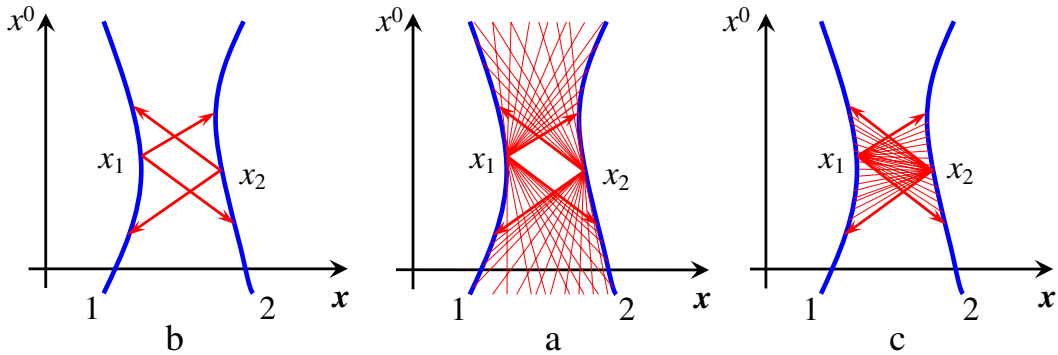}
\caption{Interaction causal structure of various Fokker-type action
integrals (specified in the text). Solid curves depict world lines of particles 1 and 2.
Arrows and thin lines depict generatrices and inwards (a) or outwards (c) of light cones
where points of particle world lines are related.}
\end{figure}

One can propose third equivalent formulation of the problem which is most convenient
for our purpose. Using the equality $\Theta(x^2)=1-\Theta(-x^2)$ one obtains:
%
%           Equation 2.7
\begin{eqnarray}\lab{2.7}
I_{\rm int}^{\rm(v)} &=& \frac a2\int^{\infty}_{-\infty}\int^{\infty}_{-\infty}
\D{}{\tau_1}\D{}{\tau_2}\,\dot x_1\!\cdot\!\dot x_2\, \left[1-\Theta(-x_{12}^2)\right]\nn\\
&=&-\frac a2\int^{\infty}_{-\infty}\int^{\infty}_{-\infty}\D{}{\tau_1}\D{}{\tau_2}\,
\dot x_1\!\cdot\!\dot x_2\,\Theta(-x_{12}^2)\\
&&{}+\frac a2 x_1\!\cdot\! x_2\Big|_{\tau_1=-\infty}^{\tau_1=\infty}\Big|_{\tau_2=-\infty}^{\tau_2=\infty}.
\nn
\end{eqnarray}
The interaction causal structure of the integral \re{2.7} is shown
in Fig.1(c). The integrals of motion are the same as in the Rivacoba
version of the model. This version (i.e., \re{2.7}) of the Weiss
action \re{2.2}, \re{2.3} is equally substantiated by the Kiskis
field theory since the function $-\Theta(-x^2)$ satisfies
the equation \re{2.5} as well.

%%%%%%%%%%%%%%%%% Section 3 %%%%%%%%%%%%%%%%%%%%%%%%%%%%%

\section{ACO approximation in the Fokker-type dynamics.}

The construction of the Hamiltonian description of Fokker-type
systems, as a step towards quantization, is a rather difficult task
which can be realized within a certain perturbation scheme. A
commonly used quasi-relativistic approximation scheme (see, for
example, \cite{GKT87}) works well if relativistic effects are weak.
But light mesons, as two-quark systems, are essentially
relativistic, and they need another approach.

Here it is used the almost-circular-orbit (ACO) approximation scheme
developed in the previous work of the author \cite{Duv12a}. The
scheme is based on the fact that all known in literature
two-particle Fokker-type systems with attractive (in some meaning)
interaction posses exact solution of the shape of concentric planar
circular particle orbits of radii $R_a(\Omega)$ dependent on an
angular velocity $\Omega$; see \cite{Sch63,A-B70,Deg71}. In
\cite{Duv12a} this is proven for a general two-particle Fokker-type
system:
%
%           Equation 3.1
\begin{eqnarray}\lab{3.1}
I &=& \sum\limits_{a=1}^{2} \int\! \D{}{t_a} L_a\left(t_a,\B x_a(t_a),\dot{\B x}_a(t_a)\right) \nn\\
&&{}+\int\!\!\int \D{}{t_1} \D{}{t_2} \Phi\left(t_1,t_2,\B x_1(t_1),\B
x_2(t_2),\dot{\B x}_1(t_1),\dot{\B x}_2(t_2)\right)
\end{eqnarray}
which is invariant under the Aristotle group (including time and
space translations and inversions, and space rotations), at least.
Manifestly covariant Fokker-type systems
\cite{W-F49,Hav71,Ker72,Riv84,Wei86,L-M12}
which by construction are
Poincar\'e-invariant (and the more Aristotle-invariant) as well as
possessing reparametrization invariance can be reduced to the form \re{3.1}
by means of the choice of the evolution parameter $\tau_a=t_a\equiv x_a^0$;
then the particle positions are $\B x_a(t_a)=\{x^i_a(t_a)\}$ ($i=1,2,3$).
The manifestly covariant Rivacoba-Weiss system \re{2.1}, \re{2.6} (or
\re{2.7}) does possesses exact circular orbit solutions
even in a strongly relativistic domain \cite{Riv84}. Thus a set of
these solutions can serve as a zero-order approximation in a
perturbative treatment of the Fokker-type dynamics.

The invariance of the action \re{3.1} with respect to time
translations and space rotations leads to an existence of the energy
an the angular momentum integrals of motion \cite{Her85}:
%
%           Equation 3.2-3
\begin{flalign}
E&=\sum\limits_{a=1}^{2}\left\{\dot{\B
x}_a\cdot\frac{\partial}{\partial \dot{\B x}_a}-1\right\}
\left(L_a+\Lambda_a\right) +
\ints\D{}{t_1}\D{}{t_2}\left\{\frac{\partial}{\partial t_1}-\frac{\partial}{\partial t_2}\right\}
\Phi, \lab{3.2}\\
\B J&= \sum\limits_{a=1}^{2}\B
x_a{\times}\frac{\partial }{\partial \dot{\B x}_a}
\left(L_a+\Lambda_a\right)\nn\\
&\qquad{}-\ha\ints\D{}{t_1}\D{}{t_2}\!\!\left\{(\B x_1+\B
x_2){\times}\frac{\partial}{\partial \B x}+ \dot{\B
x}_1{\times}\frac{\partial}{\partial \dot{\B x}_1} - \dot{\B
x}_2{\times}\frac{\partial}{\partial \dot{\B x}_2}\right\}\Phi,
\lab{3.3}
\end{flalign}
where
$$\Lambda_1=\int_{-\infty}^{\infty} \D{}{t_2}\Phi,\quad
\Lambda_2=\int_{-\infty}^{\infty} \D{}{t_1}\Phi,
\quad \ints
\equiv\displaystyle{\int_{-\infty}^{t_1}\int^{\infty}_{t_2}-\int^{\infty}_{t_1}\int_{-\infty}^{t_2}}.
$$
On the circular orbits these integrals are functions of the angular velocity:
$E_{(0)}(\Omega)$ and $\B J_{(0)}(\BOm)$, $\B J_{(0)}\|\BOm$ so that
we can get $J_{(0)}(\Omega)$ where $J_{(0)}=|\B J_{(0)}|$ and $\Omega=|\BOm|$.

Let us transit to a non-inertial reference frame which is uniformly rotating with the
angular velocity $\BOm$. This can be done via the change of variables
$\B x_a(t_a)\to\B z_a(t_a)$: $\B x_a(t_a)=\s S(t_a)\B z_a(t_a)$ where
$\s S(t)=\exp{t\s\Omega}\in \mathrm{SO}$(3) and
the skew-symmetric matrics $\s\Omega$ is dual to the vector $\BOm$.
Within this reference frame a circular motion of particles
is described by static vectors $\B R_a$ such that
$\B R_2\uparrow\downarrow\B R_1$. Then small
perturbations of circular orbits are characterized by deviation vectors
$\B\rho_a(t_a)=\B z_a(t_a)-\B R_a$.

Expanding the action \re{3.1} in powers of $\rho_a$ yields in the lowest
non-trivial order the quadratic form:
%
%           Equation 3.4
\begin{equation}\lab{3.4}
I^{(0)} = \ha\sum\limits_{kl}\intab \D{}t\D{}{t'}
\rho^k(t)D_{kl}(t-t')\rho^l(t'),
\end{equation}
where the kernel matrics $\s D(t-t')=\|D_{kl}(t-t')\|$ is invariant
under time translations
and reversion: $\s D^\mathrm{T}(t'-t)=\s D(t-t')$ (here the multi-indeces
$k,l=(a,i),(b,j)$ has been used). Corresponding equations of motion form
a time-nonlocal linear homogeneous system:
%
%           Equation 3.5
\begin{equation}\lab{3.5}
\sum\limits_l\inta \D{}{t'} D_{kl}(t-t')\rho^l(t')=0,
\end{equation}
which possesses a certain fundamental set of solutions. Among them
the exponential solutions $\rho^k(t)=e^k(\omega)\mathrm{e}^{-\im\omega t}$
are of interest. Substituting them into the system \re{3.5} yields the set
of algebraic equations:
%
%           Equation 3.6
\begin{equation}\lab{3.6}
\sum\limits_lD_{kl}(\omega)e^l(\omega)=0,
\end{equation}
which amounts the eigenvalue-eigenvector problem for the
polarization vector $e^k(\omega)$ and the frequency $\omega$. The
latter is determined by means the secular equation $\det \sf
D(\omega)=0$ in terms of the dynamical matrix $\s D(\omega)=\int\! dt\,
\s D(t){\rm e}^{\im\omega t}$. In view of time-nonlocality of the
problem \re{3.5} the matrix entries $D_{kl}(\omega)$ are, in
general, non-polynomial functions of $\omega$, and the set of
solutions of the secular equations may be infinite. Due to symmetric
properties of the dynamical matrix this set consists of duplets if
$\omega_\alpha\in\Bbb R$ or quadruplets
$\{\pm\omega_\alpha,\pm\omega^*_\alpha,\ \alpha=1,2,\dots\}$ if
$\mbox{Im}~\omega_\alpha\ne0$. In the latter case the corresponding
solution is unbounded
and cannot be described correctly within ACO approximation (where
$\rho^k$ must be small). Thus among all eigenfrequencies we
select real ones only and arrive at the following solutions of the
system \re{3.5}:
%
%           Equation 3.7
\begin{equation}\lab{3.7}
\rho^k(t)=\sum\limits_\alpha\left\{A_\alpha
e^k_\alpha(\omega_\alpha)\ {\rm e}^{-\im\omega_\alpha t}+ A^*_\alpha
\cc{e}^k_\alpha(\omega_\alpha)\ {\rm e}^{\im\omega^*_\alpha
t}\right\},
\end{equation}
where complex amplitudes $A_\alpha$ of oscillations (modes)
parameterize the phase space of the system. Only one mode $A_r$
corresponding to mutual radial particle oscillations with the
frequency $\omega_r$ is physically meaningful. Other modes are
either kinematic ones which can be reduced via redefinition of
zero-order circular orbits, or non-physical ones which reveal
physically unacceptable behavior of particles and arose as a
mathematical artefact of the theory (as in the Lorentz-Dirac
equation, for example). All such modes should be discarded.
After this is done and the polarization vectors
$e^k_\alpha(\omega_\alpha)$ in \re{3.7} are appropriately
normalized, the angular momentum and the energy of the system
take the form:
%
%           Equation 3.8-10
\begin{eqnarray}
J &=& J_{(0)}(\Omega),
\lab{3.8}\\
E &=& E_{(0)}(\Omega) + E_{(2)}(\Omega,A_r)
\lab{3.9}\\
\mbox{where}&&E_{(2)}(\Omega,A_r)=\omega_r(\Omega)|A_r|^2.
\lab{3.10}
\end{eqnarray}
Other integrals of motion following from the Poncar\'e-invariance of
the system vanish; they are the total momentum, $\B P=0$ and the
center-of-mass integral (boost), $\B K=0$. Thus the ACO
approximation brings the system into the center-of-mass reference
frame.

In order to construct the center-of-mass canonical description of
the system one should, first of all, to invert the relation \re{3.8}
with respect to $\Omega=\Omega(J)$. This permits us to obtain the
center-of-mass Hamiltonian which is nothing but the total mass of
the system:
%
%           Equation 3.11
\begin{equation}\label{3.11}
M=M_{(0)}(J)+M_{(2)}(J,|A_r|)\equiv \left\{E_{(0)}(\Omega)+
\omega_r(\Omega)|A_r|^2\right\}_{\Omega=\Omega(J)}.
\end{equation}
It is understood as a function of $J=|\B J|$ where components $J_i$
($i=1,2,3$) of the intrinsic angular momentum $\B J$ of the system
satisfy the Poisson bracket relations (PBR):
%
%           Equation 3.12
\begin{equation}\label{3.12}
\{J_i,J_j\} = {\varepsilon_{ij}}^kJ_k,
\end{equation}
and of the amplitude of interparticle radial oscillations
$A_r$ satisfying the PBR:
%
%           Equation 3.13
\begin{equation}\label{3.13}
\{A_r,A^*_r\}=-\im,\qquad \{A_r,A_r\}=\{A^*_r,A^*_r\}=0.
\end{equation}

In order to transit to an arbitrary reference frame one must
introduce canonical variables characterizing the state of the system
as a whole, for example, the total momentum $\B P$ and the
canonically conjugated CM position variable $\B Q$. Then a complete
Hamiltonian description of the system, i.e., ten canonical
generators of the Poincar\'e group, are determined in terms of $M$,
$\B J$, $\B P$ and $\B Q$ via the Bakamjian-Thomas (BT) model or
equivalent constructions \cite{B-T53,Duv89}. The quantization of BT
model is well elaborated \cite{Sok78,Pol89}.

In present work we are interested mainly in the spectrum of the mass
operator $\hat M$. It can be obtained directly from \re{3.11} by
means of the following substitution:
%
%           Equation 3.14-15
\begin{eqnarray}
&&\B J\to\hat{\B J}; \qquad A_r\to\hat A_r, \quad A^*_r\to\hat A^\dag_r; \nn\\
&&J\to\sqrt{\hat{\B J}{}^2}\to\sqrt{\ell(\ell+1)}\approx\ell+\ha, \quad \ell=0,1,...;
\label{3.14}\\
&&|A_r|^2\to\ha(\hat A_r\hat a^\dag_r+\hat a^\dag_r\hat A_r)\to
n_r+\ha, \quad n_r=0,1,...
\label{3.15}
\end{eqnarray}
Here the condition $n_r\ll \ell$ is implied, due to a perturbation
procedure.

%%%%%%%%%%%%%%%%% Section 4 %%%%%%%%%%%%%%%%%%%%%%%%%%%%%

\section{Rivacoba-Weiss model in ACO approximation.}

Let us consider a circular-orbit solution of the Rivacoba-Weiss
model. Using the action \re{2.1}, \re{2.7} for a system of two equal
particles of the mass $m_a\equiv m$ ($a=1,2$) and following the
general methodology proposed in \cite{Deg71} or \cite{Duv12a}, one
states a relation between the angular velocity $\Omega$ of a motion
of particles along circular orbits and the radius $R$ of these
orbits. It is convenient, instead of $R$, to handle with particle
velocities $v_a\equiv v=R\Omega$. Then the relation between $\Omega$
and $v$ can be determined implicitly, or parametrically, via an
auxiliary angle $\phi$. It is related with the velocity $v$ by means
of the equality:
%
%           Equation 4.1
\begin{eqnarray}\label{4.1}
f(\phi)&\equiv&\phi^2-4v^2\cos^2(\phi/2)=0\quad (0\le v<1)\quad\Longrightarrow \nn\\
&&\Longrightarrow \quad v^2=\frac{\phi^2}{2(1+\cos\phi)}
\quad\mbox{or}\quad v=\frac{\phi/2}{\cos(\phi/2)}.
\end{eqnarray}
In turns, we have for $\Omega$
%
%           Equation 4.2
\begin{eqnarray}\label{4.2}
\frac{m}{a}\Omega=\frac{\phi}{\Gamma
v^2}\left[1-\frac{(1-v^2)\phi}{f'(\phi)}\right]\equiv
f_\Omega^{\rm(v)}(\phi)
\end{eqnarray}
where $f'(\phi)\equiv \D{}{f(\phi)}/\D{}{\phi}$,
$\Gamma\equiv(1-v^2)^{-1/2}$ and the superscript "(v)" refers to the
vector interaction. Let us note that $v\in(0,1)$, $R\in(0,\infty)$
and $\Omega\in(\infty,0)$ if $\phi\in(0,\phi_1)$ where the angle
$\phi_1/2\equiv\chi_1=0.235\,\pi$
is a positive solution of the transcendental equation $\chi=\cos\chi$.

The integrals of (circular) motion $M_{(0)}$ and $J=J_{(0)}$ are
convenient to write down as follows:
%
%           Equation 4.3-4
\begin{eqnarray}
\frac{\Omega M_{(0)}}{a} &=&\frac{2\phi}{v^2}
\left[1+v^2-\frac{\phi(1+v^4\cos\phi)}{f'(\phi)}\right]\equiv
f_M^{\rm(v)}(\phi),
\label{4.3}\\
\frac{\Omega^2 J}{a}
&=&\ha f'(\phi)\equiv
f_J^{\rm(v)}(\phi).
\label{4.4}
\end{eqnarray}
They grow as $M_{(0)}\in(2m,\infty)$ and $J\in(0,\infty)$ if
$\phi\in(0,2\chi_1)$.

In order to study the system in ACO approximation we need to construct
the reduced 2$\times$2 dynamical matrix ${\cal D}^\bot$ \cite{Duv12a}
and then to calculate the frequency
$\omega_r$ of radial oscillations or, equivalently, the fraction
$\lambda=\omega_r/\Omega$, as a function of either $\Omega$, $J$, $v$ or
(which is most convenient) $\phi$. This is done in the Appendix A.

Here we are interested of an asymptotic expression for the total mass \re{3.11}
squared at $J\to\infty$. Within the perturbation procedure the inequality
$M_{(2)}\ll M_{(0)}$ is implied. Taking this into account one obtains:
%
%           Equation 4.5
\begin{eqnarray}\label{4.5}
M^2&\approx&M_{(0)}^2 + 2M_{(0)}M_{(2)}=M_{(0)}^2+2M_{(0)}\omega_r|A_r|^2 \nn\\
&=&\frac{M_{(0)}^2}{J}\left\{J + 2\frac{J\Omega}{M_{(0)}}\lambda|A_r|^2\right\}.
\end{eqnarray}
If the following limits
%
%           Equation 4.6-7
\begin{eqnarray}
k&=&\lim\limits_{J\to\infty}
\frac{M_{(0)}^2}{J}=\lim\limits_{\phi\to\phi_1}\frac{f_M^2(\phi)}{f_J(\phi)},
\label{4.6}\\
\varkappa&=&2\lim\limits_{J\to\infty}
\frac{J\Omega}{M_{(0)}}\lambda=2\lim\limits_{\phi\to\phi_1}\frac{f_J(\phi)}{f_M(\phi)}\lambda(\phi)
\label{4.7}
\end{eqnarray}
exist and are finite, the asymptotic value for the total mass squared \re{4.5} takes the form
%
%           Equation 4.8
\begin{equation}\label{4.8}
M^2\sim ka\{J+\varkappa|A_r|^2\}  \qquad\mbox{at}\quad J\to\infty
\end{equation}
and, upon quantization \re{3.12}, recovers the Regge trajectories \re{1.1}
with $\sigma=ka$. In the present case of the vector confinement model
%
%           Equation 4.9
\begin{equation}\label{4.9}
k^{\rm(v)}=8\chi_1(1+\sin\chi_1)\approx9.896, \qquad
\varkappa^{\rm(v)}=1.
\end{equation}

An asymptotic value of the daughter spacing coefficient
$\varkappa^{\rm(v)}=1$ matches well for a description of the tower
structure of meson spectra (see item 5 in Sec. 1). But the
slope coefficient $k^{\rm(v)}\approx9.896$ exceeds conventional
values $k=4\div8$ which occur in various potential models.
A plausible reason of
this disagreement in that the purely vector nature of interaction in
the model does not correspond to the actual relativistic structure of the
confinement interaction which is commonly opined as of scalar
\cite{LSG91,KvA04,Duv06} or scalar-vector
\cite{H-M02,HLS94,Hay96,LRR08,KvA84} type.

In order to confirm or challenge this assumption we construct in the
next section the scalar analogue of the Rivacoba-Weiss confinement
model.

%%%%%%%%%%%%%%%%% Section 5 %%%%%%%%%%%%%%%%%%%%%%%%%%%%%

\section{The Fokker-type action integral
with scalar linear confinement.}

Let the two-particle action to include the
free-particle terms \re{2.1} and the
Fokker-action integral of a scalar type \cite{L-M12}:
%
%           Equation 5.1
\begin{equation}\lab{5.1}
I_{\rm int}^{\rm(s)} = -\intab \D{}{\tau_1}\D{}{\tau_2}
\lr1\lr2\, G(x_{12}^2).
\end{equation}
If the function $G(x^2)$ is chosen in the form \re{2.3} it is
expected that the action \re{2.1}, \re{5.1} describes the
 scalar confinement interaction. Indeed, the action \re{5.1}
can be derived from the higher-derivative theory of scalar field
\cite{D-D04}.

In this case however one
encounters even more significant divergences as in the vector-type
model since not only the action itself and integrals of motion but
also the equations of motion are ill-posed. Fortunately, the remedy
to set the scalar model properly is the same: one replaces the
function \re{2.3} by
%
%           Equation 5.2
\begin{equation}\lab{5.2}
G(x^2)=\ha a\Theta(-x^2),\qquad a>0,
\end{equation}
which is analogous to the transition from the action \re{2.2},
\re{2.3} to \re{2.7} in the case of Weiss model. The replacement of
the function \re{2.3} by \re{5.2} in the action \re{5.1} may also be
treated as a renormalization of particle rest masses:
%
%           Equation 5.3
\begin{equation}\lab{5.3}
m_{0\,a}\to m_a=m_{0\,a}-\frac{a}4\int\nolimits_{-\infty}^{\infty}
\!\!\D{}{\tau_{\bar a}} \lr{\bar a},\qquad a=1,2,\quad \bar a=3-1,
\end{equation}
where $m_{0\,a}$ is an infinite bar mass of $a$th particle and $m_a$ is finite.

A subsequent consideration of the scalar model is similar to one in the vector case.
The system of equal rest masses is considered. Dynamical characteristics of
circular orbit solution are parameterized by the angle $\phi$. In particular,
for the angular velocity $\Omega$ one can obtain:
%
%           Equation 5.4
\begin{equation}\label{5.4}
\frac{m}{a}\Omega=\frac{\phi}{\Gamma}
\left[\frac{(1-v^2)\phi}{v^2f'(\phi)}-1\right]\equiv
f_\Omega^{\rm(s)}(\phi)
\end{equation}
where $f'(\phi)$, $v$ and $\Gamma$ as functions of $\phi$ are
defined in Sec. 4. In contrast to the vector case, here
$f_\Omega^{\rm(s)}(\phi)\to0$ if $\phi\to\phi_0\ne\phi_1$ where
$\phi_0/2\equiv\chi_0$ is a positive solution of the transcendental
equation:
%
%           Equation 5.5
\begin{equation}\label{5.5}
3\chi^2\cos\chi+2\chi^3\sin\chi-\cos^3\chi=0, \qquad \chi\in[0,\pi/4].
\end{equation}
The latter by means of the substitution $\chi=\pi/2-3\psi$ can be reduced
to the form:
%
%           Equation 5.6
\begin{equation}\label{5.6}
\psi=\frac\pi6-\frac{\sin{3\psi}}{6\cos\psi}, \qquad \psi\in[\pi/12,\pi/6]
\end{equation}
which is convenient to iterate the numerical solution: $\chi_0=0.151\pi<\chi_1$.
It is surprisingly that particle velocity $v\to0.535<1$ at $\chi\to\chi_0$ while
orbit radus $R\to\infty$. This distinguishes the scalar model from the vector one in which
$v\to1$ at $R\to\infty$. For the integrals of (circular) motion we have:
%
%           Equation 5.7-8
\begin{eqnarray}
\frac{\Omega M_{(0)}}{a}
&=&\frac{2\phi^2(1-v^2)^2}{v^2f'(\phi)}\equiv f_M^{\rm(s)}(\phi),
\label{5.7}\\
\frac{\Omega^2 J}{a}
&=&(1-v^2)\phi\equiv
f_J^{\rm(s)}(\phi).
\label{5.8}
\end{eqnarray}
It is easy to verify that $M_{(0)}\in(2m,\infty)$ and
$J\in(0,\infty)$ if $\phi\in(0,2\chi_0)$.

Using the functions \re{5.7}, \re{5.8} in eqs. \re{4.6}, \re{4.7} and
taking limits at $\phi\to\phi_0$ (instead of $\phi\to\phi_1$) yields
the slope and daughter spacing coefficients:
%
%           Equation 5.9
\begin{equation}\label{5.9}
k^{\rm(s)}=2.716, \qquad
\varkappa^{\rm(s)}=1.902.
\end{equation}
The latter is close to 2, as in the oscillator-like
and some string relativistic models of mesons
\cite{K-N73,Tak79,I-O94,Sim94}. The accidental degeneracy and thus
the tower structure of the mass spectrum is recovered approximately.
Again, the slope coefficient is not appropriate
(similarly to the vector model),
but it is considerably less than
the conventional values 4$\div$8.

The difference between the vector and scalar models suggests that
general features of the light meson spectroscopy
may be recovered (at least, asymptotically) within the Fokker-type model
with a scalar-vector confining interaction.

%%%%%%%%%%%%%%%%% Section 6 %%%%%%%%%%%%%%%%%%%%%%%%%%%%%

\section{The Fokker-type action integral
with a scalar-vector confining superposition.}

The Fokker-type system of two particles bound via superposition of scalar and
vector confining interactions is naturally defined by means of the action
\re{2.1} with
%
%           Equation 6.1
\begin{equation}\label{6.1}
I_{\rm int}^{(\xi)}=(1-\xi)I_{\rm int}^{\rm(s)}+\xi I_{\rm int}^{\rm(v)},
\end{equation}
where $I_{\rm int}^{\rm(s)}$ and $I_{\rm int}^{\rm(v)}$ are defined in eqs.
\re{5.1}, \re{5.2} and \re{2.7}, respectively, while $\xi\in[0,1]$
is a mixing parameter.

All the functions $f^{(\xi)}_\Omega(\phi)$, $f^{(\xi)}_M(\phi)$ and
$f^{(\xi)}_J(\phi)$ determining the dynamics and integrals of
circular motion of this model are superpositions of the functions
\re{4.2}-\re{4.4} and \re{5.4}, \re{5.7}, \re{5.8}:
%
%           Equation 6.2
\begin{equation}\label{6.2}
f^{(\xi)}(\phi)=(1-\xi)f^{\rm(s)}(\phi)+\xi f^{\rm(v)}(\phi).
\end{equation}
Then the function $[M_{(0)}^{(\xi)}(J)]^2$ which is a classical analogue of the
principal Regge trajectory, can be presented in the parametric form:
%
%           Equation 6.3-4
\begin{eqnarray}
\frac{\left[M_{(0)}^{(\xi)}\right]^2}{m^2}&=&
\left[\frac{f_M^{(\xi)}(\phi)}{f_\Omega^{(\xi)}(\phi)}\right]^2,
\label{6.3}\\
\frac{aJ}{m^2}&=&
\frac{f_J^{(\xi)}(\phi)}{\left[f_\Omega^{(\xi)}(\phi)\right]^2},
\qquad\qquad
{\phi\in[0,\phi_\xi],\atop\xi\in[0,1];}
\label{6.4}
\end{eqnarray}
it is shown in Fig. 2.
The maximal angle $\phi_\xi/2\equiv\chi_\xi$ is the
smallest positive root of the equation
$f^{(\xi)}_\Omega(2\chi)=0$. It grows monotonically
over the segment $\chi_\xi\in[\chi_0,\chi_1]$ if $\xi\in[0,1/2]$, and
$\chi_\xi=\chi_1$ if $\xi\in[1/2,0]$. Similarly, the maximal speed of particles
(at $R\to\infty$ when $M_{(0)}\to\infty$ and $J\to\infty$) grows monotonically,
$v\in[0.535,1]$ if $\xi\in[0,1/2]$, and
$v=1$ if $\xi\in[1/2,1]$.
\begin{figure}[th]
\begin{center}
\includegraphics[scale=1]{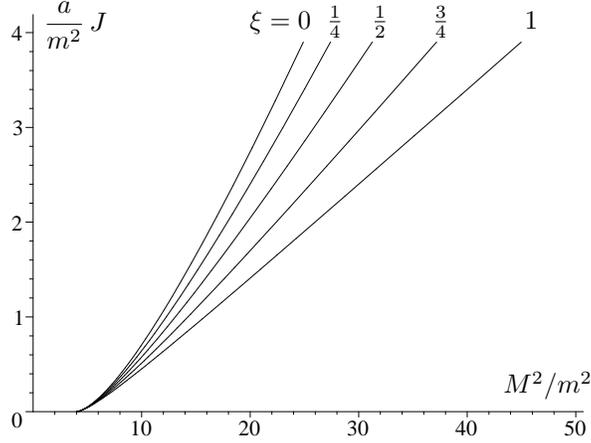}
\vspace{-1ex}
\caption{Classical Regge trajectories for different values
of the mixing parameter $\xi$.}
\end{center}
\end{figure}

The slope and daughter spacing coefficients can be calculated
similarly to the previous cases, i.e., using eqs. \re{4.6} and
\re{4.7} with the limiting angle $\phi_\xi$ (instead of $\phi_1$).
One can proof that the following equality holds:
%
%           Equation 6.5
\begin{equation}\label{6.5}
\lim\limits_{\phi\to\phi_\xi}\frac{f_J^{(\xi)}(\phi)}{f_M^{(\xi)}(\phi)}=\frac12,
\qquad \xi\in[0,1].
\end{equation}
Thus the formula \re{4.7} for the daughter spacing coefficient
simplifies:
%
%           Equation 6.6
\begin{equation}\label{6.6}
\varkappa^{(\xi)}=\lim\limits_{\phi\to\phi_\xi}\lambda^{(\xi)}(\phi), \qquad
\xi\in[0,1].
\end{equation}
The function $\lambda^{(\xi)}(\phi)$ is determined numerically from
the secular equation $\det{\bar{\cal D}^{(\xi)}(\lambda)}=0$ for the matrix
\re{A.17}; see Appendix where the graph of $\lambda^{(\xi)}(\phi)$ is presented
in Fig. 5.

Both the slope and daughter spacing coefficients are functions of the
mixing parameter. In particular,
%
%           Equation 6.7
\begin{equation}\label{6.7}
k^{(\xi)}=\xi k^{\rm(v)},\qquad\varkappa^{(\xi)}=1, \qquad
\xi\in[1/2,1].
\end{equation}
A behavior of these functions on the whole segment $\xi\in[0,1]$
is presented in Fig. 3. Grid lines on the graphs take values
of $\xi$, $k$ and $\varkappa$ into
a mutual accordance
for particular cases $\xi=1/2, 1$ and $k=4,2\pi,8$.
\begin{figure}[th]
\includegraphics[scale=1]{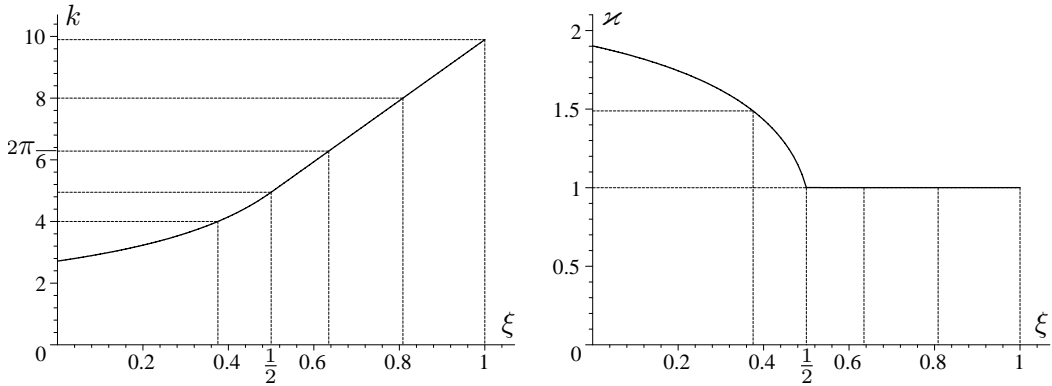}
\vspace{-1ex} \caption{Slope (left graph) and daughter spacing
(right graph) coefficients vs mix³ng parameter in the scalar-vector
model. Grid lines connect values $\xi$ with $k$ and $\varkappa$
at $\xi=1/2, 1$ and $k=4,2\pi,8$.}
\end{figure}

It is seen from these graphs that the slope coefficient
$k^{(\xi)}$ is a monotonically increasing function of the mixing parameter
$\xi$: $k^{(\xi)}\in[2.716,9.896]$ if $\xi\in[0,1]$. This segment
includes conventional values of $k=4\div8$ which occur in
non-relativistic and relativistic potential models.

Degeneracy properties of the system
with scalar-dominating confinement interaction (i.e., at
$\xi<1/2$) differ crucially from those of
$\xi>1/2$ case. In particular, the vector-dominating model
possesses the asymptotic accidental degeneracy of ($\ell{+}n_r$)-type.
Since one can provide in this case an arbitrary value for $k$ from the segment
$k^{(\xi)}\in[4.948,9.896]$,
the vector-dominating model may be compared to variety of
non-relativistic potential models and string model.

For the scalar-dominating model the lower conventional bound $k=4$
for the slope is achieved at the mixing $\xi\approx0.37$ which, in turns,
leads to the daughter spacing $\varkappa\approx3/2$.
The accidental degeneracy is present but somewhat hidden in this case.

Upon quantization of the model the mass squared spectrum is calculated by means of
the quantization rules \re{3.14}-\re{3.15} used in the classical expression \re{4.5}.
Practically, one substitutes $J=\ell+\ha$ in l.-h.s. of \re{6.4} and solves this
equation for angles $\phi_\ell$ ($\ell=0,1...$) which, in turns, are
used as arguments of the functions \re{6.3}, $f_\Omega(\phi)$ and $\lambda(\phi)$
in r.-h.s. of \re{4.5}.

\begin{figure}[h]
\begin{center}
\includegraphics[scale=1]{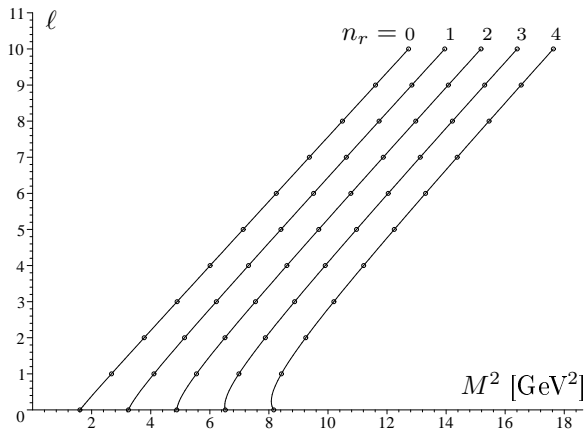}
\vspace{-1ex}
\caption{Quantum Regge trajectories for $a=0.18$~GeV$^2$, $m=0.15$~GeV and $\xi=0.63$.
The asymptotic slope $\sigma=2\pi a=1.15$~GeV$^2$,
the daughter spacing $\varkappa=1$.
}
\end{center}
\end{figure}
Let us note that classical Regge trajectories \re{6.3}, \re{6.4} (and Fig. 2) start from
$J=0$ corresponding to $\phi=0$. In the quantum case the bottom value
for the dimensionless quantity $j\equiv Ja/m^2$ (in l.-h.s. of \re{6.4})
corresponding to s-states (i.e., $\ell=0$) is $j_0\equiv \ha a/m^2>0$, hence
$\phi_0>0$. For example, taking $a=0.18$~GeV$^2$ and $m=0.15$~GeV
(the constituent mass of light quarks)
yields $j_0\approx4$. In this case a notably curved bottom segment of classical
Regge trajectories which is present in classical case (see Fig. 2)
disappears from the quantum principal trajectory
which thus is closed to a straight line \re{1.1}. Instead,
daughter trajectories acquire an erroneous curvature in their bottom, due to
an inapplicability of the quantization method at $n_r\gtrsim\ell$.
This is illustrated in Fig. 4.
It is seen an approximated tower structure of spectrum, due to
the asymptotic degeneracy of ($\ell{+}n_r$)-type.

%%%%%%%%%%%%%%%%% Section 7 %%%%%%%%%%%%%%%%%%%%%%%%%%%%%

\section{Discussion.}

In the present paper the ACO-quantization method \cite{Duv12a}
has been applied to the Rivacoba-Weiss model \cite{Riv84,Wei86}.
This model represents a Fokker-type system of two particles
which interaction can be interpreted in terms of the classical higher-derivative theory
of a vector gauge field \cite{Kis75,AAB82,A-A84,L-M12}. The Green function $\propto1/k^4$
of this field behaves as an infrared asymptotics of gluon propagator
\cite{AAB82} and leads in a nonrelativistic limit
to the linear interaction potential $U=ar$.
In the ultrarelativistic limit the model reproduces
asymptotically linear Regge trajectories whith the slope $\sigma\approx9.9a$
related rigidly to the string tension parameter $a$. The energy spectrum
reveals the accidental degeneracy of ($\ell{+}n_r$)-type which provides a tower structure
of spectrum. Thus the quantized Rivacoba-Weiss model may serve as a good base
for a description of light meson spectra.

In a variety of non-, quasi- and relativistic potential models of heavy
and light mesons the linear potential $U=ar$ appears as a scalar (or scalar-vector) long-range
part of inter-quark interaction. If one believes that the string tension
$a$ is a universal (i.e., flavor-free) parameter with conventional values in the range
$a=0.15\div0.3$~GeV$^2$ then the Rivacoba-Weiss model overestimates the slope
parameter $\sigma$. Since this model is purely vector, its counterpart based on
the higher-derivative scalar field theory \cite{D-D04} has been constructed.
The scalar model, however, underestimates the slope of Regge trajectories.
Finally, the family of scalar-vector superposition models is studied.
It turned out that the slope parameter $\sigma=1.15\div1.2\,{\rm GeV}^2$
and the string tension parameter $a=0.15\div0.3$~GeV$^2$ can be
mutually accorded if the rate of the vector interaction ranges $\xi=0.37\div0.8$.
Besides, a value of the mixing parameter $\xi$ determines
the daughter spacing parameter $\varkappa$.
In particular, $\varkappa=3/2$ at $\xi=0.37$ and $\varkappa=1$ if $\xi\ge1/2$,
so the tower structure is also provided.

It is worth to note that within non- and quasi-relativistic potential models
the linear interaction is meant mostly as a scalar one. But in many
relativistic models, especially those based on the Dirac equation,
the scalar-vector structure of a long-range interaction is preferable
\cite{H-M02,HLS94,Hay96,LRR08,KvA84}. In particular, the mixture
$\xi=1/2$, as in \cite{KvA84}, or closed values $\xi=0.48\div0.65$, as in \cite{HLS94},
enables to reduce a spin-orbital splitting in accordance to observable values.
The present relativistic model assures the scalar-vector structure
of confining interaction from another viewpoint.

In order to be appropriate for the description of
both light and heavy mesons the model should be modified.
First of all, the vector short-range interaction due to one-gluon exchange
must be introduced. It can be done naturally via complementing
the action \re{2.1}, \re{6.1} by the Wheeler-Feynman term, i.e., by \re{2.2}
with $G(x^2)=-\alpha\delta(x^2)$ where $\alpha$ is a strong coupling constant.
Then the model reproduces, in the non-relativistic limit, the Cornell potential.
This modification is expected to affect some characteristics of the model in
a relativistic regime. In particular, this may change bottom segments
of Regge trajectories and decrease their intercept $\zeta$
(see \re{1.1}) by some portion $\propto\alpha a$, similarly
to what happens in the time-asymmetric model \cite{Duv99,Duv01}.
In turns, a small intercept is appropriate for a description
of lightest mesons \cite{Tak79}. A study of the model complemented with
the Wheeler-Feynman term is beyond the scope of this work.

Another extension of the model for a sterling meson spectroscopy
is the insertion of particle spins. One can exploit, as a guideline,
a description of spinning particles in terms of anti-commuting variables
used in the Wheeler-Feynman electrodynamics \cite{KvA86}.
A quantization method should be modified appropriately.

%%%%%%%%%%%%%%%%%%%%%%%%%%%%%%%%%%%%%%%%%%%%%%%%%%%%%%%%
\section*{Acknowledgment}
The author is grateful to
V. Tretyak and Yu.~Yaremko for helpful discussion of this work.

%%%%%%%%%%%%%%%%%%%%%% Appendix %%%%%%%%%%%%%%%%%%%%%%%%%

\section*{Appendix. Calculation of ${\cal D}^\bot$ and $\lambda=\omega_r/\Omega$}
\renewcommand{\theequation}{A.\arabic{equation}}
\setcounter{equation}{0}

It is convenient to define a dimensionless 2$\times$2 reduced
dynamical matrix
%
%           Equation A.1
\begin{equation}\label{A.1}
\bar{\cal D}\equiv\frac1{a\Omega}{\cal D}^\bot
=\frac{m}{a}\Omega{\cal C} + {\cal K} - {\mit\Xi}
=f_\Omega(\phi){\cal C} + {\cal K} - {\mit\Xi}
\end{equation}
where:
%
%           Equation A.2-4
\begin{eqnarray}
{\cal C}&=&\left[\begin{array}{cc}
\Gamma^3+\lambda\Gamma & -\im\lambda\Gamma^3v^2 \\
\im\lambda\Gamma^3v^2 & \Gamma+\lambda\Gamma^3
\end{array}\right],
\lab{A.2}\\
{\cal K}&=&\int\limits_0^\phi\D{}\varphi{{\cal K}}_0 -
\left.\frac1{f'(\varphi)}{{\cal K}}_1\right|_{\varphi=\phi} +
\left.\frac1{f'(\varphi)}\frac{\D{}{}}{\D{}{\varphi}}
\frac1{f'(\varphi)}{{\cal K}}_2\right|_{\varphi=\phi},
\lab{A.3}\\
{\mit\Xi}&=&\int\limits_0^\phi\D{}\varphi{{\mit\Xi}}_0 -
\left.\frac1{f'(\varphi)}{{\mit\Xi}}_1\right|_{\varphi=\phi} +
\left.\frac1{f'(\varphi)}\frac{\D{}{}}{\D{}{\varphi}}
\frac1{f'(\varphi)}{{\mit\Xi}}_2\right|_{\varphi=\phi}. \lab{A.4}
\end{eqnarray}
The matrix $\cal C$ comes from the free-particle term of the action
\re{2.1}. The function $f_\Omega(\phi)$ and components of other
matrices ${\cal K}$ and ${\mit\Xi}$ depend on the interaction model.

For the vector (Rivacoba-Weiss) model the function
$f_\Omega^{\rm(v)}(\phi)$ in defined in \re{4.2}, and matrices in
r.h.s. of \re{A.3} and \re{A.4} have the form:
%
%           Equation A.5-10
\begin{eqnarray}
{{\cal K}}^{\rm(v)}_0&=&0,
\lab{A.5}\\
{{\cal K}}^{\rm(v)}_1&=& 2 \left[\begin{array}{cc}
1{+}v^2\mathrm{c}(3{+}2\mathrm{c}) & 0 \\
0 & 2v^2\mathrm{s}^2
\end{array}\right]
-2\im\lambda v^2(1+\mathrm{c}) \left[\begin{array}{cc}
0 & 1 \\
-1 & 0
\end{array}\right],
\lab{A.6}\\
{{\cal K}}^{\rm(v)}_2&=& -4v^2(1+v^2\mathrm{c})
\left[\begin{array}{cc}
(1{+}\mathrm{c})^2 & 0 \\
0 & \mathrm{s}^2
\end{array}\right],
\lab{A.7}\\
{{\mit\Xi}}^{\rm(v)}_0&=& (1+\lambda^2) \left[\begin{array}{cc}
\mathrm{c}\mathrm{C} & \mathrm{s}\mathrm{S} \\
-\mathrm{s}\mathrm{S} & \mathrm{c}\mathrm{C}
\end{array}\right]
-2\im\lambda \left[\begin{array}{cc}
\mathrm{s}\mathrm{S} & -\mathrm{c}\mathrm{C} \\
\mathrm{c}\mathrm{C} & \mathrm{s}\mathrm{S}
\end{array}\right],
\lab{A.8}\\
{{\mit\Xi}}^{\rm(v)}_1&=& -2 \left[\begin{array}{cc}
\mathrm{c}(1{+}v^2(2{+}3\mathrm{c}))\mathrm{C} & \mathrm{s}(1{+}v^2(1{+}3\mathrm{c}))\mathrm{S} \\
-\mathrm{s}(1{+}v^2(1{+}3\mathrm{c}))\mathrm{S} &
(1{+}v^2(\mathrm{c}^2{-}2\mathrm{s}^2))\mathrm{C}
\end{array}\right]
\nn\\
&&{}-2\im\lambda v^2 \left[\begin{array}{cc}
2\mathrm{s}(1{+}\mathrm{c})\mathrm{S} & (\mathrm{s}^2{-}\mathrm{c}(1{+}\mathrm{c}))\mathrm{C} \\
(\mathrm{c}(1{+}\mathrm{c}){-}\mathrm{s}^2)\mathrm{C} &
2\mathrm{s}\mathrm{c}\mathrm{S}
\end{array}\right],
\lab{A.9}\\
{{\mit\Xi}}^{\rm(v)}_2&=& 4v^2(1+v^2\mathrm{c})
\left[\begin{array}{cc}
(1{+}\mathrm{c})^2\mathrm{C} & \mathrm{s}(1{+}\mathrm{c})\mathrm{S} \\
-\mathrm{s}(1{+}\mathrm{c})\mathrm{S} & -\mathrm{s}^2\mathrm{C}
\end{array}\right],
\lab{A.10}
\end{eqnarray}
where $\mathrm{s}\equiv\sin\varphi$, $\mathrm{c}\equiv\cos\varphi$,
$\mathrm{S}\equiv\sin(\lambda\varphi)$,
$\mathrm{C}\equiv\cos(\lambda\varphi)$.

For the scalar confining interaction the function
$f_\Omega^{\rm(s)}(\phi)$ is defined in \re{5.4}, and matrices in
r.h.s. of \re{A.3} and \re{A.4} have the form:
%
%           Equation A.11-16
\begin{eqnarray}
{{\cal K}}^{\rm(s)}_0&=& \left[\begin{array}{cc}
\Gamma^2 & 0 \\
0 & 1
\end{array}\right]
+-\im\lambda(\Gamma^2+1) \left[\begin{array}{cc}
0 & 1 \\
-1 & 0
\end{array}\right]
+\lambda^2 \left[\begin{array}{cc}
1 & 0 \\
0 & 1
\end{array}\right],
\lab{A.11}\\
{{\cal K}}^{\rm(s)}_1&=& 2 \left[\begin{array}{cc}
1{-}v^2(3{+}2\mathrm{c}) & 0 \\
0 & 1-v^2
\end{array}\right]
-2\im\lambda v^2(1+\mathrm{c}) \left[\begin{array}{cc}
0 & 1 \\
-1 & 0
\end{array}\right],
\lab{A.12}\\
{{\cal K}}^{\rm(s)}_2&=& -4v^2(1-v^2) \left[\begin{array}{cc}
(1{+}\mathrm{c})^2 & 0 \\
0 & \mathrm{s}^2
\end{array}\right],
\lab{A.13}\\
{{\mit\Xi}}^{\rm(s)}_0&=& \Gamma^2v^2\mathrm{C}
\left[\begin{array}{cc}
1 & -\im\lambda \\
\im\lambda & \lambda^2
\end{array}\right]
\lab{A.14}\\
{{\mit\Xi}}^{\rm(s)}_1&=& -2 \left[\begin{array}{cc}
(\mathrm{c}(1{-}3v^2){-}2v^2)\mathrm{C} & \im(1{-}2v^2)\mathrm{s}\mathrm{S} \\
-\im(1{-}2v^2)\mathrm{s}\mathrm{S} & (1{-}v^2)\mathrm{c}\mathrm{C}
\end{array}\right]
\nn\\
&&{}-2\im\lambda v^2 \left[\begin{array}{cc}
0 & (1{+}\mathrm{c})\mathrm{C} \\
-(1{+}\mathrm{c})\mathrm{C} & -2\mathrm{s}\mathrm{S}
\end{array}\right],
\lab{A.15}\\
{{\mit\Xi}}^{\rm(s)}_2&=& 4v^2(1-v^2) \left[\begin{array}{cc}
(1{+}\mathrm{c})^2\mathrm{C} & \mathrm{s}(1{+}\mathrm{c})\mathrm{S} \\
-\mathrm{s}(1{+}\mathrm{c})\mathrm{S} & -\mathrm{s}^2\mathrm{C}
\end{array}\right].
\lab{A.16}
\end{eqnarray}
\begin{figure}[h]
\begin{center}
\includegraphics[scale=1]{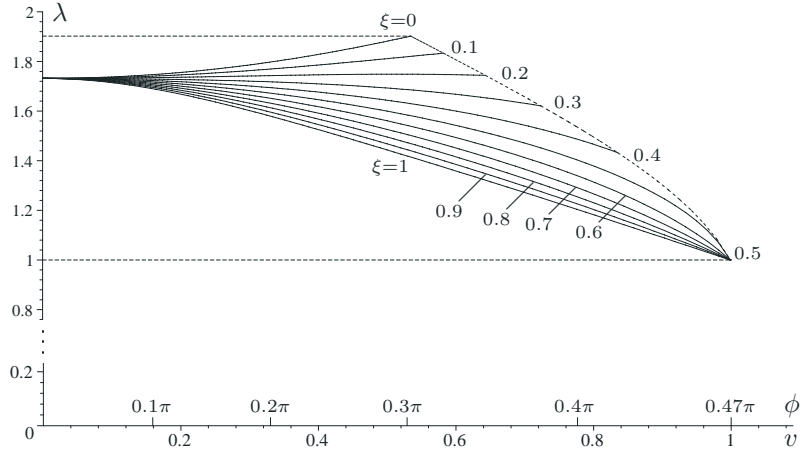} \vspace{-4ex}
\end{center}
\caption{The relative frequency $\lambda=\omega_r/\Omega$ as a
function of the angle $\phi$ and the velocity $v$ of particle
circular motion for different values of the mixing parameter $\xi$.}
\end{figure}

For the scalar-vector superposition the dimensionless dynamical
matrix is constructed as follows:
%
%           Equation A.17
\begin{equation}\label{A.17}
\bar{\cal D}^{(\xi)}=(1-\xi)\bar{\cal D}^{\rm(s)}+\xi\bar{\cal
D}^{\rm(v)},
\end{equation}
where $\xi$ is the mixing parameter. The relative frequency
$\lambda$ is then calculated as a real positive root of the reduced
secular equation $\det{\bar{\cal D}(\lambda)}=0$. In general, this
can be done numerically.

In Fig. 5 the relative frequency $\lambda=\omega_r/\Omega$ as a
function of the velocity $v$ of particle circular motion is shown
for various values of the mixing parameter $\xi$. Let us note that
$$
\lim\limits_{v\to0}\frac{\omega_r}{\Omega}= \sqrt{3}
$$
as it must be for the nonrelativistic problem with the linear
potential $U=ar$ \cite{Duv12a}.

%%%%%%%%%%%%%%%%%%%%%% References %%%%%%%%%%%%%%%%%%%%%%%%%

\end{document}